\documentclass[aip,preprint]{revtex4-1}
\usepackage{dcolumn}% Align table columns on decimal point
\usepackage{graphicx}% Include figure files
\graphicspath{{./fig/}}
\usepackage{bm}% bold math
\usepackage[T1]{fontenc}
\usepackage{mathptmx}
\usepackage{here}

\draft % marks overfull lines with a black rule on the right

\begin{document}

\title{Characterization of a double torsion pendulum used to detect spin-induced torque based on Beth's experiment} %Title of paper

 \affiliation{Department of Applied Physics, Tokyo University of Agriculture and Technology, Koganei,
Tokyo 184-8588, Japan%\\This line break forced with \textbackslash\textbackslash
}%
\author{Runa Yasuda}

 %\altaffiliation[]{Department of Applied Physics, Tokyo University of Agriculture and Technology.}%Lines break automatically or can be forced with \\
\author{Atsushi Hatakeyama}%
 \email{hatakeya@cc.tuat.ac.jp}

\date{\today}
\begin{abstract}
We characterized a double torsion pendulum system, including measurements of the photon-spin-induced torque. Our experimental strategy was similar to that used in Beth's experiment, which was performed in 1936 to measure photon-spin-induced torque using forced oscillation caused by polarization modulation of light incident on a suspended object. Through simple passive isolation of the suspended object from external vibration noise, the achieved torque sensitivity was $2 \times 10^{-17}$~N m in a measurement time of 10$^4$ s, which is close to the thermal noise limit and one order smaller than the minimum torque measured in Beth's experiment. The observed spin-induced torque exerted on the light-absorbing optics is consistent with the angular momentum transfer of $\hbar$ per photon.

\end{abstract}

\maketitle %\maketitle must follow title, authors, abstract and \pacs

\section{\label{sec:Intro}Introduction}
The torsion pendulum, or torsion balance, is a sensitive device used to detect weak forces. \cite{Gil93} It was used in the experiments of Cavendish and Coulomb, conducted in the 18th century to measure gravitational and electrostatic constants. The torsion pendulum is still one of the most sensitive measurement methods in modern physics and engineering, with applications ranging from low-frequency gravitational wave detection \cite{Shi20} to studies of the mechanical properties of materials. \cite{Kes21}

The torsion pendulum has been used to detect small mechanical torques exerted on macroscopic objects from quantum mechanical spins. The first experiment was performed in 1915 by Einstein and de Haas, \cite{Ein15} who demonstrated that the mechanical rotation of a ferromagnetic material is caused by flipping of its magnetization or internal atomic spins. The second experiment, by Beth, was carried out in 1936 \cite{Beth36} and demonstrated that flipping the photon spin from $+\hbar$ to $-\hbar$ ($\hbar$: Planck's constant divided by $2\pi$) induced the rotation of a transmitting half-wave plate. The results of both experiments are understood in terms of angular momentum conservation of the system of atomic or photonic spins, and the rotation of the macroscopic object. 
Spin-induced torque has received renewed attention in recent spintronics and optomechanics studies. Angular momentum transmission due to a spin wave has been mechanically detected with an yttrium ion garnet cantilever. \cite{Har19} Gigahertz rotation of optically levitated nanoparticles has been demonstrated via spin transfer arising from circularly polarized trapping light. \cite{Rei18, Jon18}

In our previous report, \cite{Hat19} we discussed the development of a torsion pendulum to study spin transfer from an optically spin-polarized atomic gas to a solid gas container through atom-surface interactions. The experiment was similar to that of Beth, in which an object suspended with a thin wire was irradiated from the bottom with circularly polarized light. Beth's experiment has been described in many textbooks. \cite{Hec, Joa} However, to our knowledge, the careful and patient measurements conducted therein have not been replicated using the original forced oscillation method, partly due to experimental difficulties including radiation pressure torque and the so-called radiometer effects, as well as the multiple steps involved in deriving the torque from the measured amplitude of the forced oscillation, as pointed out in Ref.~\cite{Del05,Emi18}. In Ref.~\cite{Del05}, the authors directly measured photon-spin-induced torque on the order of $10^{-12}$~N m from the acceleration of a suspended half-wave plate with a high-power infrared laser.

In this study, we analyzed our developed pendulum system, including measurement of the photon-spin-induced torque. Our experimental strategy was similar to the strategy applied in Beth's experiment: forced torsional oscillation induced by periodic application of external torque at the resonance frequency of the pendulum. The main differences between our system and that of Beth's study are the use of a double pendulum, in which the first pendulum works as a vibration isolator, and a higher operating frequency of 0.1~Hz relative to 0.002~Hz in Beth's experiment.
Through this simple passive isolation from external vibration noise, the torque sensitivity was $2 \times 10^{-17}$~N m (in 10$^4$ -s measurement time), which is close to the thermal noise limit and one order smaller than the minimum torque measured in Beth's experiment. The observed spin-induced torque exerted on the light-absorbing optics is consistent with the angular momentum transfer of $\hbar$ in the direction of light propagation per photon for left-circularly polarized light. \cite{Hec,Cor}
Therefore, we concluded that our torsion pendulum properly detected the photon-spin-induced torque.

This paper is organized as follows. We first explain the experimental apparatus, with special attention given to the design of the double pendulum in Sec.~\ref{sec:App}. In Sec.~\ref{sec:Exp}, we describe the experimental procedure, including a forced torsional oscillation approach and a lock-in data analysis method. We then present our experimental results in Sec.~\ref{sec:Res}, in which measurement of photon-spin-induced torque and estimation of the torque sensitivity  are discussed. The paper is concluded in Sec.~\ref{sec:Con}.
 
\section{\label{sec:App}Experimental apparatus}

\begin{figure}
\includegraphics[width=85mm]{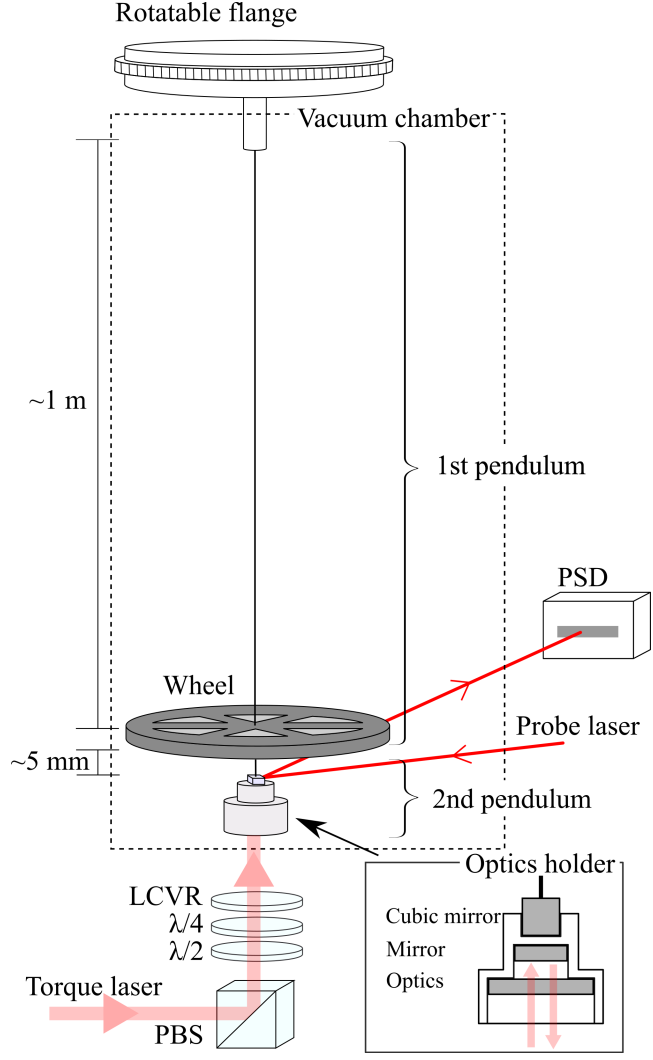}
\caption{\label{app}Schematic diagram of the experimental apparatus. PBS: polarization beam splitter; PSD: position-sensitive detector; $\lambda/4$: quarter-wave plate; $\lambda/2$: half-wave plate; LCVR: liquid crystal variable retarder.}
\end{figure}

The experimental apparatus, shown in Fig.~\ref{app}, is a slightly modified version of our previously reported one. \cite{Hat19}
The double torsion pendulum, discussed in detail below, was suspended in a vacuum chamber fixed on an optical table. The pressure inside was maintained at $1\times10^{-5}$~Pa with an ion pump, which was lower than the radiation pressure exerted on a suspended object by a typical torque laser beam used in the experiment ($1 \times10^{-4}$~Pa for a laser intensity of $3\times10^4~\rm{W/m^2}$). Thus, the radiometer effect from the background gas was negligible. 

The angle of the suspended object (``optics holder'') was monitored using an optical lever method with a red probe laser (CPS650F; Thorlabs Inc.) and a position-sensitive detector (PSD) (S3932; Hamamatsu Photonics K.K.) located $L=0.38$~m from the torsional axis. The angle displacement $d\phi$ was derived from the position displacement $dx$ measured with the PSD as $d\phi=dx/(2L)$.
Light from the torque laser, a volume-holographic-grating-stabilized laser with a wavelength of 852~nm (LD852-SEV600; Thorlabs Inc.), was incident on the object from the bottom. The light polarization was controlled with a set of waveplates, including a liquid crystal variable retarder (LCVR) (LCC1223-B; Thorlabs Inc.), to periodically flip the polarization of the torque laser.
Signal processing and data recording were systematically performed with PC-based digital and analog input/output devices controlled by LabVIEW programs (NI).

Compared to Beth's experiment, which used a quartz fiber 0.25 m in length, the optics holder to be rotated was suspended with a short tungsten (W) wire approximately 5~mm in length and 10~$\mu$m in diameter. We refer to this portion of our double pendulum as the second pendulum. 
The short wire resulted in a higher torsional resonance frequency of 0.1 Hz, compared to 0.002 Hz in Beth's experiment. To compensate for this, the second pendulum was mounted at the end of another pendulum, which had a resonance frequency of 0.0008 Hz to provide better isolation from external vibration noise, as discussed below.
The higher resonance frequency also provided shorter measurement times, which minimized spurious effects of drifting. The optics holder accommodated half-inch optics, on which torque was exerted from the torque laser beam; a mirror reflected the torque laser beam when it transmitted the optics. A cubic mirror was attached to the top of the optics holder for the optical lever. The moment of inertia of the optics holder was estimated to be $(9.4 \pm 0.9)\times10^{-8}$~kg m$^2$ from the drawing. The uncertainty originated from inaccuracy in the manufacturing process; it was estimated by comparing the designed and measured weights of the optics holder. 

The optics holder suspended with the short wire was attached to another torsion pendulum (the first pendulum), which consisted of an object (the ``wheel'') with a large moment of inertia and a long wire. The moment of inertia of the wheel was estimated to be $7\times 10^{-6}$~kg m$^2$, and the W wire was approximately 1~m in length and 10~$\mu$m in diameter. The first pendulum had a low resonance frequency of torsional oscillation at 0.0008~Hz and served as a vibration isolator for the second pendulum, as discussed below.

\begin{figure}[h]
\includegraphics[width=85mm]{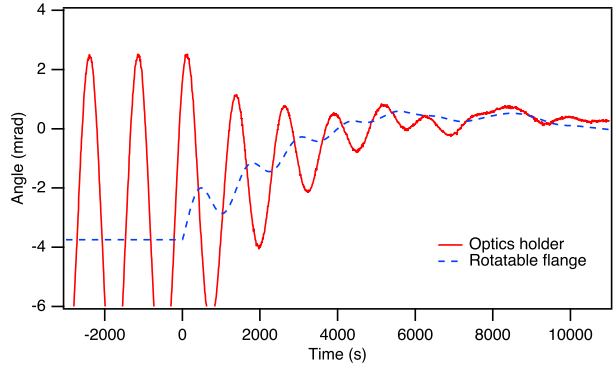}
\caption{\label{damp}Damping of the torsional oscillation of the first pendulum with the rotatable flange. The angle of the rotatable flange was controlled electronically, while the angle of the optics holder was measured with the optical lever. The curve representing the angle of the optics holder was low-pass-filtered to clarify the oscillation of the first pendulum at its resonance period (1240~s). At 0 s, the automatic proportional integral control to the flange angle was activated toward the target angle of 0 mrad for the optics holder.}
\end{figure}

The long wire was attached to a vacuum chamber flange (VRS-70M; Shinku-Kogaku Corp.), which was rotatable with an electric stepper motor (0.00004$^{\circ}$/pulse). This rotatable flange was used to dampen the torsional oscillation of the first pendulum. Figure \ref{damp} shows the angle of the rotatable flange controlled electronically, as well as the angle of the optics holder measured with the optical lever. The curve representing the angle of the optics holder was low-pass-filtered to clarify the oscillation of the first pendulum at its resonance period (1240~s). At 0~s in Fig. \ref{damp}, proportional integral control to the flange angle was activated toward the target angle of 0~mrad for the optics holder to dampen the resonance oscillation of the first pendulum. This damping technique was quite useful for readying the system for forced oscillation measurements for the second pendulum, after the apparatus experienced large external perturbations such as earthquakes, which often caused large and slowly damping torsional oscillation of the first pendulum.

\begin{figure}[h]
\includegraphics[width=85mm]{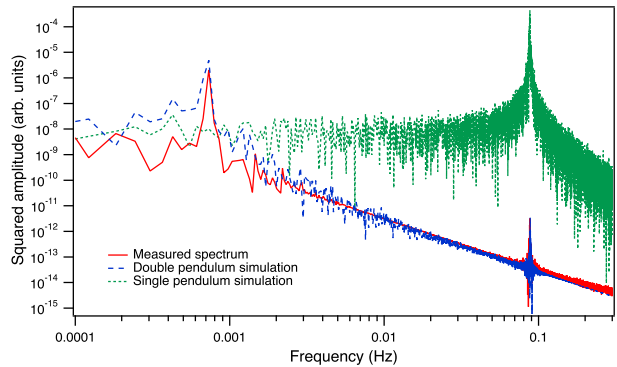}
\caption{\label{sim}Fourier transform spectrum of free torsional oscillation of the pendulum for $2\times10^4$-s measurement. Numerical simulations for the double and single pendulums are also shown.}
\end{figure}

Figure \ref{sim} shows the Fourier transform spectrum of a free oscillation, recorded for $2\times 10^4$~s after damping of the first pendulum oscillation had been completed. The peak at 0.0008~Hz corresponds to the resonant torsional oscillation of the first pendulum. The amplitude decreased as the frequency increased beyond the resonance frequency. The peak at 0.1~Hz corresponds to the torsional oscillation of the second pendulum and is the focus of this study.

To quantitatively examine the effect of vibration isolation with the first pendulum, we numerically solved the coupled equations that describe the torsional motions of the two pendulums with application of white noise between 0 and 2~Hz to the fixed end of the first pendulum wire; the freedom of motion in the simulation was the rotation of the first and second pendulums around the fixed wire axis. With appropriate experimental parameters, as shown in Fig.~\ref{sim}, the Fourier transform spectrum of the simulation reproduced the experimental observations fairly well, particularly the decrease in baseline oscillation amplitude at higher frequencies than the resonance of the first pendulum. The effect of vibration isolation can be clearly understood by comparing the measured spectrum with the single pendulum simulation performed on the assumption that the second pendulum was attached directly to the flange without the first pendulum. 

%However, the peak for the resonant torsional oscillation of the second pendulum was higher than that expected from the double pendulum simulation. We presumed that the external vibration noise that induced the vertical, horizontal, and tilt motions of the wire-fixed point was coupled to the torsional motion of the second pendulum. \cite{Gil93} The high amplitude of the measured spectrum at lower frequencies can be attributed to larger external vibration noise at low frequencies.

\section{\label{sec:Exp}Experiments}
We adopted a forced oscillation method at the resonance frequency of the second pendulum. 
A numerical simulation using the coupled equations for the double pendulum confirmed that the second pendulum could be regarded as a single pendulum around its resonance frequency, while the first pendulum (to which the second pendulum was attached) oscillated at a much lower resonance frequency. Therefore, we can use basic knowledge regarding the forced oscillation of the single pendulum, as described below.

In a steady state, the frequency dependence of the quadrature component of the forced oscillation with respect to the periodic external torque $N'\cos\omega t$ ($\omega$: angular frequency; $t$: time) corresponded to the following Lorentzian function around the resonance frequency:
\begin{equation}
\frac{\mu N'}{4I\omega_R}\frac{1}{(\omega-\omega_R)^2+(\mu/2)^2}, \label{quadrature}
\end{equation}
where $I$ is the moment of inertia, $\omega_R$ is the resonance angular frequency, and $\mu$ is the full width at half maximum of the resonance, determined by friction proportional to the angular velocity. Therefore, the peak amplitude of the resonance $A$ is $N'/(\mu I\omega_R)$.
The torque modulation $N'$ was then derived from the quadrature component of the torsional oscillation at the resonance, as follows: 
\begin{equation}
N'=\mu I \omega_R A=\frac{I\omega_R^2A}{Q},\label{N’}
\end{equation}
where $Q=\omega_R/\mu$ is the Q-factor of the resonance. 

In this study, the optical torque dependent on the photon spin was modulated in a square wave by changing the torque laser polarization with the LCVR, rather than in a sinusoidal wave. According to the Fourier expansion, $\pm N$ square modulation in the torque corresponds to $N'=\frac{4}{\pi} N$ sinusoidal modulation. Therefore, the factor $4/\pi$ was considered when deriving the torque modulation $N$ from the measured quadrature component as follows:
\begin{equation}
N=\frac{\pi}{4}\mu I \omega_R A=\frac{\pi}{4}\frac{I\omega_R^2A}{Q}.\label{N}
\end{equation}
We evaluated $\omega_R$ and $Q$ in every set of measurements to derive $N$ from the measured $A$.

The quadrature and in-phase components of the oscillation (i.e., the oscillation amplitude and phase delay with respect to the reference signal, which was synchronized with the modulation of the torque laser polarization) were derived by lock-in analysis of the forced oscillation using Igor Pro data analysis software (WaveMetrics, Inc.).

The spin-induced torque modulated by flipping the polarization of the torque laser between the left- and right- circular polarizations is ${P}_{\rm{laser}}\lambda/(2\pi c)= {P}_{\rm{laser}}[W]\times 4.52 \times 10^{-16}$~N m, where ${P}_{\rm{laser}}$ and $\lambda$ are the power and the wavelength of the torque laser, respectively, and $c$ is the speed of light, because one photon carries $+\hbar$ or $-\hbar$ spin component along the laser propagation. 
Special precautions were required against torque caused by radiation pressure. 
The torque exerted on the optics by the radiation pressure of the torque laser beam $\bm{N}_{\rm{rad}}$ can be calculated by the following equation:
\begin{equation}
\bm{N}_{\rm{rad}}= \bm{r} \times \bm{F}_{\rm{rad}}, 
\end{equation}
where $\bm{r}$ is the position at the laser beam enters the optics with respect to the wire-fixed position, and $\bm{F}_{\rm{rad}}$ is the radiation force.
If the torque laser beam is incident either perfectly parallel to or onto the wire-fixed point, the torque around the rotation axis is zero. Assuming imperfect incidence of the torque laser with a deviation of $a$ and an incline of $\alpha$ from the rotation axis, the torque around the rotation axis is $\alpha |\bm{F}_{\rm{rad}}| a=\alpha P_{\rm{laser}} a/c$.
When we measure optical torque in polarization modulation mode, the ratio of modulation of the radiation pressure torque to that of the spin-induced torque is $2\pi(a/\lambda)\alpha\beta$, where $\beta$ is a factor representing the modulation of the radiation pressure torque. This modulation may have been caused by variation in the laser power. Careful adjustment of the optical system to minimize the imperfect incidence and power variation associated with the polarization modulation is required to make the factor $2\pi(a/\lambda)\alpha\beta$ much less than $1$, such that the spin-induced torque modulation dominated the radiation pressure torque modulation. It was not difficult to fulfill this requirement, assuming $a=1$~mm, $\alpha=10^{-2}$~rad, and $\beta=10^{-3}$ for $\lambda=852$~nm. The effect of radiation pressure was examined by intentionally modulating the power of the torque laser in the experiment.

\section{\label{sec:Res}Results}
Because the optics holder was able to accommodate various types of half-inch optics, we examined several optics. 
We mainly focused on a neutral density (ND) filter (NE560B-B; Thorlabs Inc.) as a light absorber, because our future target, a solid container of an atomic gas, can be regarded as a light absorber. \cite{Hat19} The absorption of the torque light by the ND filter was greater than 99.9\%. The spatial mode of the laser beam was filtered to be TEM$_{00}$ with a single-mode fiber. The laser power incident on the ND filter was 132~mW. The polarization of the torque laser was modulated between left-circular and right-circular polarization using the LCVR.

\begin{figure}[h]
\includegraphics[width=85mm]{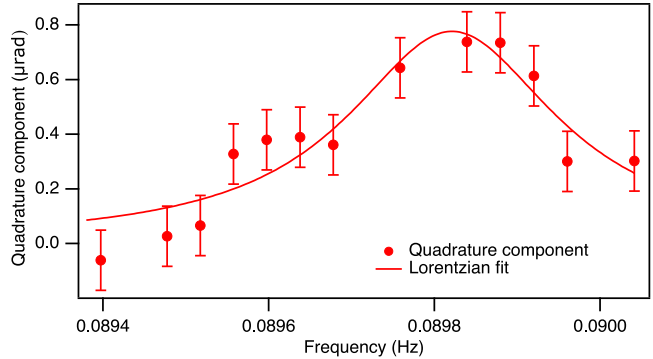}
\caption{\label{ND_spc}Resonance spectrum of the second pendulum with the ND filter irradiated with counterrotating circularly polarized light. Each error bar was estimated to be $\pm0.1~\mu$rad (see the text).}
\end{figure}

Figure \ref{ND_spc} shows the quadrature component of the induced torsional oscillation as a function of the modulation frequency around the resonance frequency of the second pendulum. One data point was the average of data collected for $10^4$~s in a steady state. The estimation of the error bar is discussed below. Lorentzian fitting yielded a peak amplitude of $0.78 \pm 0.08$~$\mu$rad at the resonance frequency of $0.08982 \pm 0.00002$~Hz, and a full width at half maximum value of $(3.1 \pm 0.5) \times 10^{-4}$~Hz. The Q-factor was derived as $(2.9 \pm 0.5)\times 10^2$. Therefore, the measured torque modulation was $(6.3 \pm 1.3)\times 10^{-17}$~N m, derived using Eq.~(\ref{N}). The systematic uncertainty, mainly originating from conversion of the position signal measured with the PSD to the angle of the optics holder, was less than 5\%, smaller than this statistical uncertainty of $1.3\times 10^{-17}$~N m. The measured torque was consistent with the theoretically estimated value of $(6.0 \pm 0.2) \times 10^{-17}$~N~m derived from the assumption that one photon transfers $\hbar$ angular momentum to the ND filter; the main origin of the uncertainty in the theoretical estimation originated from the laser power.

\begin{figure}[h]
\includegraphics[width=85mm]{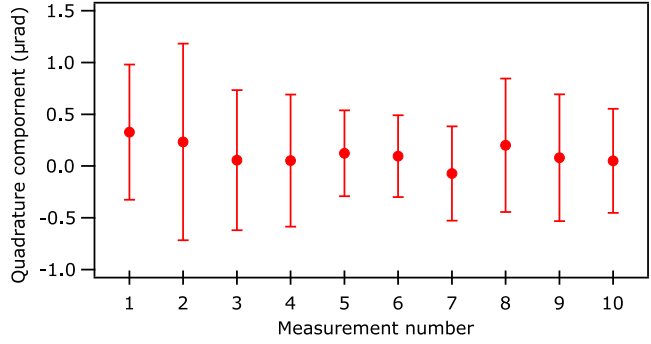}
\caption{\label{Noise} Fluctuation of quadrature components of the free torsional oscillation. Each error bar represents the standard deviation for each $10^4$-s measurement.}
\end{figure}

The error bars in Fig. \ref{ND_spc} were estimated from different measurements. We obtained 10 sets of $10^4$-s free torsional oscillation data for the pendulum, and applied lock-in analysis to them with imaginary reference signals at the resonance frequency. Figure~\ref{Noise} shows the obtained quadrature components of the oscillation with the standard deviations. Each set of data had a standard deviation similar to the case of the measurements in Fig. \ref{ND_spc}. The 10 measurements resulted in an average quadrature component of $0.11~\rm{\mu rad}$ and a standard deviation of $0.11~\rm{\mu rad}$. We therefore estimated that the uncertainty in detection of the quadrature component was $0.1~\rm{\mu rad}$ in a single $10^4$-s measurement; we adopted this value as the error bar for each data point shown in Fig. \ref{ND_spc}.

For the second pendulum used to estimate the uncertainty mentioned above, we found that $\omega_R=0.110317\pm0.00007$~Hz and $Q=173\pm5$. Using Eq.~(\ref{N}) with these parameters, the uncertainty in the quadrature component detection of $0.1~\rm{\mu rad}$ was converted to the uncertainty in the torque detection of $2 \times10^{-17}$~N m. We regarded this as the torque sensitivity of our system over a measurement time of 10$^4$ s.
The thermal noise limit of the torque $N_{\rm{ther}}$ can be estimated using the following formula, \cite{Hai07}
\begin{equation}
N_{\rm{ther}}=\frac{\pi}{4}\sqrt{\frac{4k_BTI\omega_R}{Q\Delta t}} \label{min_torque},
\end{equation}
where $k_B$ is the Boltzmann constant, $T$ is temperature, and $\Delta t$ is the measurement time. The estimated thermal noise limit was $2 \times10^{-17}$~N m, which is close to the experimentally estimated sensitivity.

 We also confirmed that the observed oscillation phase was consistent with the expected rotation direction; specifically, left-circularly polarized light carries $+\hbar$ angular momentum along the direction of propagation \cite{Hec,Cor} and exerts a positive torque on the ND filter.

\begin{figure}[h]
\includegraphics[width=85mm]{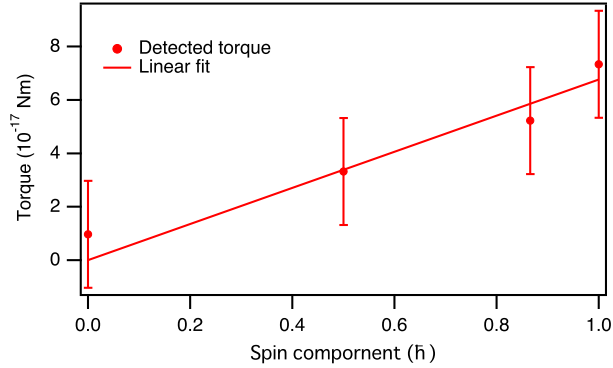}
\caption{\label{ND_po} Polarization dependence of the photon-spin-induced torque. Each error bar was estimated to be $\pm2 \times 10^{-17}$~N~m (see the text).}
\end{figure}

To confirm that the observed torque originated from photon spin, rather than from other effects such as radiation pressure torque, we measured the light polarization dependence. At a fixed modulation frequency, we changed the polarization modulation by rotating the LCVR axis from a pair of counterrotating circular polarizations to a pair of counterrotating elliptical polarizations. Figure \ref{ND_po} shows the measured torque as a function of the spin component along the light propagation direction. The spin component of $0~\hbar$ corresponds to nominally unmodulated linear polarization, whereas the spin component of $1~\hbar$ corresponds to circular polarizations, with elliptical polarizations between the two extremes. The observed torque is reasonably proportional to the photon spin component of the propagation direction.

We also modulated the power of the torque laser without changing the polarization, to estimate the torque modulation originating from the radiation pressure; specifically, a $\pm$7\% change in power induced a torque modulation of $1 \times10^{-16}$~N~m. The power modulation associated with the polarization change with the LCVR was 0.1\%. Therefore, we estimated that the torque modulation originating from the radiation pressure was $1\times10^{-18}$~N~m, which was negligible in the observed torque modulation.
Taken together, we concluded that the observed torque was induced by photon spin transfer.

\section{\label{sec:Con}Conclusions}
In this paper, we describe the characterization of our developed double torsion pendulum system using optical torque with a forced oscillation method similar to Beth's experiment. In the double pendulum approach, the first pendulum operates as a vibration isolator. With this simple passive isolation from external vibration noise, the torque sensitivity reached $2 \times 10^{-17}$~N m in 10$^4$ -s measurement time, which is close to the thermal noise limit and one order smaller than the minimum torque measured in Beth's experiment.
We evaluated the photon-spin-induced torque using a ND filter as a light absorber. The observed torque was consistent with the torque expected from the angular momentum transfer of $\hbar$ (or $-\hbar$) per photon in left- (or right-) circularly polarized light. We concluded that our torsion pendulum properly detected the photon-spin-induced torque. Our demonstration can be regarded as a recapitulation of Beth's experiment.

Finally, the developed double pendulum could be used in future detailed studies of spin transfer from gases to solids.

\begin{acknowledgments}
We thank Jean Jeener for his useful comments.
This study was supported by the Matsuo Foundation.
\end{acknowledgments}

\section*{Data Availability Statement}
The data that support the findings of this study are available from the corresponding author upon reasonable request.

\providecommand{\noopsort}[1]{}\providecommand{\singleletter}[1]{#1}%

\end{document}